\documentstyle[11pt,aaspp4]{article}

\def\beq{\begin{equation}}
\def\eeq{\end{equation}}
\def\beqar{\begin{eqnarray}}
\def\eeqar{\end{eqnarray}}

\def\pref#1{(\ref{#1})}
\def\pcite#1{(\cite{#1})}

\def\la{\mathrel{\mathpalette\fun <}}

\def\fun#1#2{\lower3.6pt\vbox{\baselineskip0pt\lineskip.9pt
  \ialign{$\mathsurround=0pt#1\hfil##\hfil$\crcr#2\crcr\sim\crcr}}}

\def\gyr{\,{\rm Gyr}}
\def\myr{\,{\rm Myr}}
\def\kyr{\,{\rm kyr}}
\def\yr{\,{\rm yr}}

\def\psec{\,{\rm pc}}

\def\gram{\,{\rm g}}
\def\bp{\,{\sc bp}}
\def\mm_myr{\, {\rm mm} \, {\rm Myr}^{-1}}
\def\t12{t_{1/2}}
\def\msol{\hbox{$M_{\odot}$}}

\def\iso#1#2{\hbox{${}^{#2}${\rm #1}}}
\def\be1#1{\iso{Be}{1#1}}
\def\mn5#1{\iso{Mn}{5#1}}
\def\fe5#1{\iso{Fe}{5#1}}
\def\fe6#1{\iso{Fe}{6#1}}
\def\kr8#1{\iso{Kr}{8#1}}
\def\i12#1{\iso{I}{12#1}}
\def\sm14#1{\iso{Sm}{14#1}}

\def\tsn{t_{\rm SN}}

\begin{document}

\begin{flushright}
CERN-TH/98-373 \\
astro-ph/9811457 \\
\end{flushright}

\title{ON DEEP-OCEAN ${}^{\bf 60}$Fe AS A FOSSIL OF A NEAR-EARTH
SUPERNOVA}

\author{Brian D. Fields}
\affil{Department of Astronomy, University of Illinois \\
Urbana, IL 61801}

\author{John Ellis}
\affil{Theoretical Physics Division, CERN \\
Geneva, Switzerland}
            
\begin{abstract} 
Live \fe60 has recently been reported
in a deep-ocean ferromanganese crust.  Analysis of the isotopic ratios
in the sample suggests that the measured \fe60 abundance
exceeds the levels generated by terrestrial and cosmogenic
sources, and it has been proposed that the 
excess of \fe60 is a signature of
a supernova that exploded near the earth several $\myr$ ago.  
In this paper, we consider the possible background
sources, and confirm that the measured \fe60 is significantly higher
than all known backgrounds, in 
contrast with the reported abundance of live \mn53 in the same sample.
We discuss scenarios in which the data are consistent with a supernova event 
at a distance $D \sim 30 \psec$ and an epoch $\tsn \sim 5 \myr$ ago.
We  propose tests that could confirm or refute the interpretation of the 
\fe60 discovery, including searches for \be10, \i129 and \sm146. Such a 
nearby supernova event
might have had some impact on the earth's biosphere, principally
by enhancing the cosmic-ray flux. This might have damaged the earth's
ozone layer, enhancing the penetration of solar ultraviolet radiation.
In this connection, we comment on the Middle Miocene and Pliocene 
mini-extinction events. We also speculate on the possibility of
a supernova-induced ``cosmic-ray winter'',
if cosmic rays play a significant role in seeding cloud
formation.
\end{abstract}


\section{Introduction}

The literature contains considerable discussion of the
likelihood of nearby supernova events: their
frequency has been estimated (Shklovsky \cite{shkl}), and their
possible impacts on the biosphere have been considered
(Ruderman \cite{rude}; Ellis \& Schramm \cite{es};
Ellis, Fields, \& Schramm \cite{efs}). Within the
considerable uncertainties, it is conceivable that there
may have been one or more nearby supernova events during the Phanerozoic era.
This has prompted discussion of their isotopic signatures and kill radii,
and speculation on their possible role in triggering biological
mass extinctions.

Supernova events that were sufficiently close to have left some
terrestrial isotope signature, but far enough not to have triggered
a mass extinction, are expected to have been more frequent.
In this connection, various authors have noted the enhancement of
\be10 in ice cores and
marine sediments $\sim 35 \kyr \bp$ ($\bp =$ before the present).
In particular, Ellis, Fields, \& Schramm \pcite{efs} discussed the 
possibility that this might have arisen from the supernova event that
gave birth to the Geminga pulsar, and
proposed looking at deep-ocean sediments, suggesting that the
long-lived isotopes \i129, \sm146, and \iso{Pu}{244},
as well as the shorter-lived \be10, \iso{Al}{26}, \iso{Cl}{36}, \mn53, 
\fe60, and \iso{Ni}{59}, might provide geological evidence of a nearby
supernova event at any time during the past $10^8$~yr or more.

In the light of this proposal, the
recent deep-ocean ferromanganese crust measurements (Knie et al.\
\cite{fe60})
of \fe60 ($\t12 = 1.5 \myr$) and \mn53 ($\t12 = 3.7 \myr$)
are very exciting.  Whilst \mn53 is known to have a significant
natural background from interplanetary dust accretion,
the expected background for \fe60 is
significantly lower than the measured levels.
Knie et al.\ discuss possible alternative origins
for the apparent excess of \fe60, and argue that their results can only be
understood in terms of a supernova origin
for the \fe60.  As we discuss in more detail later, their
data suggest a nearby supernova event within $\sim 30$ \psec ~during the
past few $\myr$, a significantly longer timescale than 
had been discussed in connection with the \be10 signal.  

The data are very new and the statistics is small,
with only 63 \fe60 
and \mn53 nuclei being detected in total.
However, if these data are confirmed, together with their extraterrestrial
interpretation, the implications are profound.  
The would constitute the first direct evidence that a supernova 
event occurred near earth within a relatively
recent geological time, with detectable effects on our planet. The purpose
of this paper is to discuss the implications of the data in more detail.

We first re-estimate the cosmogenic backgrounds to both the \fe60 and
\mn53 signals observed by Knie et al.\ \pcite{fe60}. 
We confirm their estimates that
the \fe60 signal is higher than plausible cosmogenic sources, whereas
the \mn53 signal is compatible with such backgrounds. We emphasize the
desirability of understanding better the sedimentation history of the
last $20$~Myr, and seeking confirmation in other ferromanganese crusts
and elsewhere that the apparent \fe60 excess is global. 
We stress in particular the desirability of finding an earlier
layer in which the apparent \fe60 excess is absent. 
We then use the available
data to derive constraints on the possible supernova event, including its
time and distance. We also discuss the possible implications for
supernova nucleosynthesis and review other isotope signals - such as
\be10, \i129 and \sm146 - that could be
used to confirm the supernova diagnosis and deepen its
interpretation.  Finally, we address the possible implications
of this apparent
supernova event for the terrestrial biosphere. We revisit the
effects of the expected cosmic-ray flux on the earth's ozone layer,
and the resulting enhanced penetration of the atmosphere by solar
ultraviolet radiation. We speculate whether its effects could be related
to either of the mini-extinctions that apparently occurred during the
Middle Miocene and Pliocene epochs. We also raise the possibility of
other supernova-induced effects on the biosphere, in particular the
possibility that the enhanced cosmic-ray flux might seed extra
cloud cover,  potentially triggering global cooling: a ``cosmic-ray
winter''.

\section{Data}

Knie et al.\ \pcite{fe60} studied the isotopic composition of a deep ocean
sample of hydrogenic ferromanganese crust.
This is material which has slowly
precipitated from seawater onto
one of several specific kinds of substrate.
Using accelerator mass spectrometry,
Knie et al.\ detected both live
\fe60 and live \mn53 in 
three layers at different depths, spanning 0 to 20 mm.  They
estimate that these depths correspond to times
spanning 0 to 13.4 Myr $\bp$. Their data are summarized 
below. From the outset of our analysis, we emphasize caution, since the
data may yet turn out to be a false alarm. 
We re-analyze the consistency of the data with conventional
cosmogenic backgrounds before exploring the 
consequences if their
interpretation in terms of a nearby supernova explosion is confirmed.
We emphasize the necessity for follow-up data to
confirm or reject this tantalizing scenario.

\begin{table}[htb]
\caption{
\protect{Ferromanganese crust data from Knie et al.\ \pcite{fe60}}.} 
\begin{tabular}{|cc|ccc|ccc|}
\hline\hline
depth & age $\Delta t_\ell$ & \mn53 & $\phi_{53}(\Delta t_\ell)$ 
   & $N_{53}(\Delta t_\ell)$ & \fe60 & $\phi_{60}(\Delta t_\ell)$ 
   & $N_{60}(\Delta t_\ell)$ \\
mm & \myr \bp& events & ${\rm cm}^{-2} \myr^{-1}$ 
   & ${\rm cm}^{-2}$ & events & ${\rm cm}^{-2} \myr^{-1}$ 
   & ${\rm cm}^{-2}$ \\
\hline\hline
$0-3$ & $0-2.8$ & 26 & $2.6 \times 10^{8}$ & $5.7 \times 10^{8}$ 
   & 14 & $1 \times 10^{6}$ & $1.6 \times 10^{6}$ \\
$5-10$ & $3.7-5.9$ & 6 & $6.4 \times 10^{8}$ & $5.8 \times 10^{8}$ 
   & 7 & $7 \times 10^{6}$ & $1.8 \times 10^{6}$ \\
$10-20$ & $5.9-13.4$ & 7 & $4.4 \times 10^{8}$ & $6.0 \times 10^{8}$ 
   & 2 & $9 \times 10^{6}$ & $1.3 \times 10^{6}$ \\
\hline\hline
\end{tabular}
\label{tab:data}
\end{table}

Knie et al.\ quantify their detection in terms of the
flux $\phi_i(\Delta t_\ell)$
deposited in layer $\ell$ of the crust, which they infer as follows.
Using an independently determined crustal growth rate
(which varies from $1-2 \mm_myr$),
Knie et al.\ translate the depth of each layer 
into time intervals $\Delta t_\ell = (t_{\ell,{\rm i}},t_{\ell,{\rm f}})$
before the present.  
The fluxes are calculated given the number of detected events in each
layer, corrected for radioactive decays, 
and assuming a constant deposition during each time interval.
The reported fluxes appear in Table \ref{tab:data}.
They found that the inferred fluxes of \fe60 increase as one
goes back in time, whereas the inferred fluxes of \mn53
are less variable.
Since both radioisotopes are found in all three layers,
whereas a nearby supernova would naively 
be expected to have contaminated at most one 
layer, one might conclude that the background must be significant
for both isotopes. However, as we discuss below, astrophysical
effects might have spread out the deposition of the signal,
and one should not ignore the possibility of some terrestrial 
mixing effect such as bioturbation. 
Knie et al.\ estimate that the \mn53 background is large and do not
use this as the primary indicator of a supernova signal.
In \S \ref{sec:bgnd},
we explicitly calculate the expected background levels.

In our analysis below, we find it useful to express
the observations in terms of
the (present-day, uncorrected) 
surface density $N_i^{\rm obs}(\Delta t_\ell)$ detected in 
crust layer $\ell$.
To extract this from the reported fluxes one
simply inverts the procedure used to derive the fluxes
\beqar
N_i^{\rm obs}(\Delta t_\ell) 
  & = & 
    \int_{t_{\ell,{\rm i}}}^{t_{\ell,{\rm f}}} \ dt \ \phi_i(t)
    \ e^{-t/\tau_i}  \\
  & = & 
    \left[ \exp\left({-\frac{t_{\ell,{\rm i}}}{\tau_i}}\right)
        -  \exp\left({-\frac{t_{\ell,{\rm f}}}{\tau_i}}\right) \right]
    \  \tau_i
    \  \phi_i(\Delta t_\ell)
\eeqar
The inferred surface densities also appear in Table \ref{tab:data}.
Note that, although the inferred flux is strongly varying with time,
the present surface density is quite constant.
We examine later reasons why the \fe60 signal
is seen in multiple crust layers, in apparent conflict with the
simplest picture of punctuated deposition of material after
a nearby supernova explosion.

As noted by Knie et al., the observed flux of material into the
ferromanganese crust is not necessarily the same as the mean flux of
material averaged over the globe.  The largest difference between
the two arises from the 
reduced uptake of both Mn and Fe into the crust, i.e.,
the efficiency for each of these elements to
be deposited onto the crust is not perfect,
and thus the flux onto the crusts is lower than the
flux deposited over the earth's surface.  
Consequently, the surface densities in the crust and globally are 
related by 
\beq
\label{eq:reduced}
N_i^{\rm obs} = f_i \ N_i^{\oplus}
\eeq
where $N_i^{\oplus}$ is the global surface density deposited,
and $f_i < 1$ accounts for the reduced uptake.
Thus, to compare with the observations, one must 
reduce the theoretically
expected surface density, $N_i^{\oplus}$, by
the uptake factor $f_i$ in order to compare with the
observations.
In our calculations, we use the fiducial values suggested
by Knie et al., who recommend approximate values of
$f_{53} \sim 1/20$, and $f_{60} \sim f_{53}/5 \sim 1/100$, which
are based in part on estimates of the \mn53 background.
We find that these estimates have significant uncertainties, which
translate directly into corresponding uncertainties in $f_{53}$, 
$f_{60}$, and finally $N_{60}^{\oplus}$.

\section{Interpreting \fe60 in the Nearby Supernova Hypothesis}

Ellis, Fields \& Schramm \pcite{efs} estimated the terrestrial deposition
of
observable long-lived $\beta$-unstable nuclei.
Since the deposition rate is a few $\mm_myr$, any 
radioisotopic signature should be
temporally isolated, unless there is some astrophysical effect or
disturbance of the deposited layers
that could smear out the time resolution.  However, even if that is the
case,
one can still constrain the deposited fluence $F_i$ of the radioisotopes
and the time since deposition.  
If one measures a surface density $N_i$ of atoms per unit area,
which was deposited at a time $t$ before the present, one knows that
then 
\beq
\label{eq:pres_dens}
N_i (t) = \frac{1}{4} \ F_i \ e^{-t/\tau_i}
\eeq
where we assume that the fall-out on the earth's surface is isotropic,
the factor of $1/4$ is the ratio of the area of the earth's shadow
($\pi R_\oplus^2$) to the surface area of the earth ($4 \pi R_\oplus^2$),
and $\tau_i$ is the mean life of $i$.
If one assumes an isotropic ejection of supernova debris,
the deposited fluence is in turn
directly related to the supernova distance $D$ and the yield:
\beq
\label{eq:dep_flue}
F_i = \frac{M_i/A_i m_p}{4 \pi D^2} 
\eeq
where the mass ejected in species $i$ is $M_i$.  
Note that we have assumed for simplicity that
essentially no decays occur as the ejecta is transported to the
earth, i.e., that the transit time $\delta t \ll \tau_i$.
As we expect
$\delta t \la few \times 10^5 \yr$ for a supernova blast, 
this is an excellent approximation
within the accuracy of our calculation.
Combining \pref{eq:pres_dens} and \pref{eq:dep_flue}, we have 
\beqar
\label{eq:master}
N_i (t) 
   & = & \frac{M_i/A_i m_p}{16 \pi D^2} \ e^{-t/\tau_i} \\
\nonumber
   & = & 4.6 \times 10^{8} \ e^{-t/\tau_i} \ {\rm cm}^{-2} \
      \left( \frac{A_i}{60} \right)^{-1} \;
      \left( \frac{M_i}{10^{-5} \msol} \right) \;
      \left( \frac{D}{30 \psec} \right)^{-2}
\eeqar
We re-emphasize that the surface density calculated here
assumes (1) an isotropic explosion, (2) isotropic fall-out
on the earth's surface, and (3) that the
incorporation of the debris into the crust or sediment that
is sampled is faithful (i.e., no chemical fractionation
and 100\% uptake).  The real situation is likely to violate
all of these assumptions at some level.
In particular, Knie et al.\ note that fractionation and
reduced uptake effects
have already been observed and are large.  Such effects
must be taken into account before one can use \pref{eq:master}
to compare theory and observation.

If one can indeed deduce the supernova contribution to the
abundance of {\em one} radioisotope in a sediment or crust,
then one can determine the supernova distance with the addition of
two other inputs.  That is, with the observations
$N_i^{\rm obs}$ in hand,
one can solve \pref{eq:master} 
for $D$, given estimates of the yield $M_i$ and
the deposition epoch $\tsn$, as determined by the depth of the 
supernova layer in the crust:
\beq
\label{eq:D_SN}
D = e^{-\tsn/2 \tau_i} \ \sqrt{ \frac{f_i M_i}{16 \pi A_i m_p N_i^{\rm obs}} }
\eeq
where we have allowed for the reduced uptake 
as in \pref{eq:reduced}.
If one can deduce the supernova depositions of {\em two} 
radioisotopes $i$ and $j$ (where $\tau_i < \tau_j$), 
then one can use \pref{eq:master} with the combination 
the yields and the observed $N_i$ and $N_j$ to get not only
$D$ but also an independent estimate of $t$, as follows:
\beq
\label{eq:t_SN}
\tsn = \frac{1}{ \tau_i^{-1} - \tau_j^{-1} } \
   \ln \left( \frac{A_j}{A_i} \, \frac{f_i}{f_j} \, 
       \frac{N_j^{\rm obs}}{N_i^{\rm obs}} \, \frac{M_i}{M_j} \right) 
\eeq
This value is derived independently of the value of
$t$ inferred from the depth
of the supernova-enhanced layer in the crust.
The two values can thus be compared 
as a consistency check.

For the estimates below, we adopt the supernova yields from
Woosley \& Weaver \pcite{ww},
one of the most extensive studies to date of
supernova nucleosynthesis.
Woosley \& Weaver tabulate isotopic yields
for nuclei with $A<66$ over
a range in progenitor
masses and progenitor metallicity. Since the putative
supernova would have occurred close to the Sun, we
adopt the yields for solar metallicity.
We note that, whilst the \mn53 yield is expected to grow as a
fairly smooth function of the progenitor mass,
the \fe60 yields are not.  Consequently, the
\mn53/\fe60 ratio varies widely and non-monotonically.
The maximum is $\mn53/\fe60 \simeq 20$ at $20 \msol$,
and the minimum is $\mn53/\fe60 \simeq 0.6$ at $13 \msol$.
This introduces an additional uncertainty,
since the mass of the nearby supernova is unknown.
The initial mass function
favors stars in the $11-20 \msol$ range, 
whose $\mn53/\fe60$ ratios span this range. We
adopt the fiducial yield $M_{60} = 10^{-5} \msol$,
corresponding to an intermediate value, recognizing that this is
clearly somewhat uncertain.

\subsection{Backgrounds}  
\label{sec:bgnd}

There are several potential contributions to the background.
(1) Cosmogenic production, namely the
spallative  production of radioisotopes due
to the steady cosmic-ray flux into the
earth's atmosphere, was discussed in detail in Ellis, Fields \& Schramm.  
We have repeated that analysis using the cross section for 
\fe60 production via spallation in $p+\kr84$, as
reported in Knie et al.\ \pcite{fe60}.  We agree with these authors
that the \fe60 from this source is several orders of magnitude
below the observations.
(2) A related source is {\it in situ} production, via 
the penetrating muon and neutron flux.  We have estimated
thus using the calculations by Lal \& Peters \pcite{lp},
and find that this source is negligibly small.
(3) In the absence of a strong cosmogenic component, the
dominant contribution to the background of
both elements is 
the influx of extraterrestrial material, e.g., dust and meteorites, onto the 
earth. This material has been exposed to cosmic rays en route, and thus
also contains the resulting spallogenic species at some level.

To compute the radioisotope contribution of infalling
material, one must first determine the present rate $J$ of meteoric
mass accretion onto the earth.
This quantity is difficult to measure, and past estimates
have spanned a full six orders of magnitude, as seen in
the tabulation of Peucker \& Ehrenbrink (\cite{p-e}).
However, recent measurements using different
techniques have showed a convergence.
Love \& Brownlee \pcite{lb} have
inferred the micrometeor ($\la 10^{-4} \gram$) 
mass spectrum and flux from the cratering patterns
on the exposed surfaces of the Long Duration Exposure Facility satellite,
and have reported a total accretion rate of
$J = (4 \pm 2) \times 10^{10} \gram \yr^{-1}$
due to micrometeor infall.
Love \& Brownlee report that these objects have a
mass spectrum which peaks around $1.5 \times 10^{-5} \gram$,
corresponding to a diameter $\sim 200 \, \mu {\rm m}$,
and have an infall rate which probably dominates
the mass accretion on short timescales,
with large (multi-ton) impacts possibly contributing
a similar net influx on longer timescales.
An independent measure of extraterrestrial
dust accretion comes from the analysis of osmium concentrations
and isotopic ratios in oceanic sediments.
Recent inferred accretion rates of $(3.7\pm 1.3) \times 10^{10} \gram
\yr^{-1}$
from Peucker \& Ehrenbrink (\cite{p-e}), and of 
$(4.9-5.6)\times 10^{10} \gram \yr^{-1}$  from Esser \& Turekian
\pcite{et}
agree with each other and with the Love \& Brownlee result.
We adopt a fiducial rate of $J = 4 \times 10^{10} \gram \yr^{-1}$,
which we expect to be accurate to within 50\%.

With the meteor accretion rate in hand,
we now estimate the cosmogenic background by adopting the following
simplified picture. 
In general, some fraction of the infalling material
does not impact the earth directly as
a meteorite, but is mixed into the atmosphere.
This ``well-mixed'' fraction
includes all of the smallest bodies (micrometeorites),
and the portion of the larger bodies which is vaporized
during the descent\footnote
{The vaporized material comprises the outermost layers of
the falling material. For the larger objects these are also the regions
with
the highest concentration of spallogenic material.}.   
Since the accretion rate includes
micrometeorites and dust, we assume that all such material is
indeed well-mixed. We further assume that
it is deposited isotropically on the surface of the earth. 
The total isotropic mass flux of incoming material is thus
\beq
\dot{\Sigma} = \frac{J}{4 \pi R_\oplus^2} 
\eeq
If the average mass fraction of radioisotope $i$
in the well-mixed infalling material
is $X_i$,
then the isotropic mass flux of $i$ is just 
$\dot{\Sigma}_i = X_i \dot{\Sigma}$,
and the number flux of $i$ is
\beqar
\nonumber
\Phi_i  & = & \frac{X_i \dot{\Sigma}}{A_i m_p}  \\
\label{eq:meteor}
        & = & 4.7 \times 10^{21} \ \frac{X_i}{A_i} 
              \ {\rm atoms} \ {\rm cm}^{-2} \ \myr^{-1}
              \ \left( \frac{J}{4 \times 10^{10} \ \gram \, \yr^{-1}} \right)
\eeqar
The problem now reduces to finding $X_i$.

For \mn53, we use the results of Michel et al.\ \pcite{mich},
who calculate the production of cosmogenic nuclides
in meteoroids by cosmic-ray protons.  These calculations
are tabulated in Michel et al.\ \pcite{mich} for different kinds of 
meteoroids as functions of depth and size. We use
the zero-depth values, since these correspond
to the smallest and most common objects.
Michel et al.\ express their results in terms
of the specific activity $\Gamma_i$, 
i.e., the decay rate per unit mass of iron.
Finally, we take the infalling material to have
the iron mass fraction $X_{\rm Fe} = 0.19$,
as found in C1 carbonaceous chondrites (Anders \& Grevesse \cite{ag}).
Then, in a meteorite with such a mass faction  of
iron, we have $X_i = m_i \tau_i \Gamma_i X_{\rm Fe}$,
and so
\beq
X_{53} =  1.9 \times 10^{-11} \ 
   \left( \frac{X_{\rm Fe}}{0.19} \right) \
   \left( \frac{\Gamma_{53}}{400 \, {\rm dpm} \, {\rm kg \, Fe}^{-1}} \right)
\eeq
Using this, we estimate a background flux of \mn53 of
\beq
\Phi_{53} =  1.7 \times 10^{9}
   \ {\rm atoms} \ {\rm cm}^{-2} \ \myr^{-1}
\eeq
For comparison, Bibron et al.\ \pcite{bib} report a measurement of
\mn53 in antarctic snow which implies
\beq
\Phi_{53}^{\rm obs} =  (6.1 \pm 1.4) \times 10^{9} 
   \ {\rm atoms} \ {\rm cm}^{-2} \ \myr^{-1}
\eeq
whereas Imamura et al.\ \pcite{imam} measured \mn53 in ocean sediments and
found
\beq
\Phi_{53}^{\rm obs} =  (2.0 \pm 0.9) \times 10^{9} 
   \ {\rm atoms} \ {\rm cm}^{-2} \ \myr^{-1}
\eeq
The differences in these results highlight the large systematic errors
in these estimates. However,
within these uncertainties, we find our estimate to be 
in good agreement with the data.  

The \fe60  background
is lower because \fe60 lies above the iron peak,
thus reducing the abundance of the possible target nuclei. 
Because of this and the shorter lifetime
of \fe60, the data on
\fe60 in meteors are much sparser, and we know of just
two relevant results in the literature.
Early work by Goel \& Honda \pcite{gh}
detected \fe60 in the Odessa meteorite
with a specific activity
of $\Gamma_{60} = 0.9 \pm 0.2 \, {\rm dpm} \, {\rm kg}^{-1}$,
for a sample with 7\% Ni and 91\% Fe,
or $\Gamma_{60}^{\rm Ni} = 13 \pm 3 \, {\rm dpm} \, {\rm kg \, Ni}^{-1}$.
This report can be used to make a background estimate, as we did for
\mn53.  Using the specific activity of Goel \& Honda 
and a Ni mass fraction $X_{\rm Ni} = 0.011$, we would find an \fe60 flux of 
\beq
\Phi_{60} =  1.3 \times 10^{6}
   \ {\rm atoms} \ {\rm cm}^{-2} \ \myr^{-1}
\eeq
This can be compared with the value we obtain from the
\fe60/\mn53 ratio reported by Knie et al. for the Dermbach meteorite,
which implies that 
$\Gamma_{60}^{\rm Ni} = 1.2 \times 10^{-2} \Gamma_{53}^{\rm Fe}$.
For our adopted $\Gamma_{53}^{\rm Fe}$, this gives
$\Gamma_{60}^{\rm Ni} = 5  \, {\rm dpm} \, {\rm kg \, Ni}^{-1}$,
and 
\beq
\Phi_{60} =  0.5 \times 10^{6}
   \ {\rm atoms} \ {\rm cm}^{-2} \ \myr^{-1}
\eeq
In order to compare with the levels measured by Knie et al.\
in the ferromanganese crust, one must reduce  
the total \fe60
flux by the uptake factor $f_{60} \sim 1/100$ discussed earlier
(\ref{eq:reduced}).  Thus the expected background \fe60 is
at least two orders of magnitude smaller that the observed level in the
crust.

Knie et al.\ argue, however, that the \mn53 and \fe60 
interstellar cosmogenic production mechanisms differ significantly.
Namely, they argue that while \mn53 comes primarily from spallation
of Fe nuclei by high-energy Galactic cosmic rays, \fe60 derives
mostly from reactions on Ni induced by secondary neutrons.
These secondary neutrons would be more abundant in the interiors of
large meteors than in micrometeors or interstellar dust.
In the absence of secondary neutrons, the \fe60 production
by protons alone is much smaller, leading to a lower
specific activity $\Gamma_{60}$ and finally a lower flux on earth.
This scenario can be tested experimentally, by measuring the
\fe60 depth profile in meteorites.  

An effect omitted from this simple calculation of
the meteoritic background 
is that, if a nearby supernova occurs, it leads to an
enhanced cosmic-ray flux not only on the earth, 
but also in the entire solar system, including, e.g., 
the material which falls as meteorites.
Thus, one expects the meteoritic ``background'' to
in fact include also some supernova ``signal,'' and thus to 
undergo an increase which lasts for
a timescale of order the species' lifetime.
For the present data, this is not
a serious issue.  The enhanced cosmic-ray 
flux lasts for at most a few kyr,
whereas the meteoritic sources average over several Myr, so the
increase in the measured fluence for any given layer is
only perceptible if the cosmic-ray flux enhancement is 
much larger than expected: the needed
increase is a factor $\sim 10^3$, which is
about an order of magnitude more than expected.
However, if the signal could be measured with a much finer
time resolution of order $10-100$ kyr, then this effect could
be significant.
At any rate, this discussion points up the advantage of
measuring \mn53 content in layers
{\em prior} to the alleged supernova event, which should not show
any supernova-related enchancement.

\subsection{Distance and Epoch of the Putative Supernova}

A crucial problem for any attempt to estimate these parameters
is that the observed signals are present in all layers for both isotopes.
We assume that the \fe60 signal is real and not due to
background, but not necessarily the \mn53 signal.  Even so, any deposition 
mechanism should be able to accommodate
the continuous, rather than punctuated and isolated,
nature of the signal.

We thus consider two possible scenarios.
\begin{enumerate}

\item As suggested 
by Knie et al., a continuous \fe60 signal could arise from
residual contamination due to the nearby supernova.  Knie et al.
note that this could contaminate the local 
interstellar medium, in particular
contaminating dust in the local interstellar medium which could enter the
solar cavity and fall onto the earth.
Also, the cosmic-ray flux would irradiate metoric and cometary
material in the solar cavity, leading to enhanced \fe60 and \mn53
production.  Both processes lead to a \fe60 flux which has
an abrupt onset but is continuous until the \fe60 is extinct.
Until it is extinct, the present-day \fe60 levels would be constant,
assuming a constant dust accretion rate, since all of the \fe60 was 
created at the same time, by the 
supernova event.  This scenario is compatible with the observations.

\item Alternatively, it is possible that the \fe60 signal
is punctuated but mixed in the sample itself, 
e.g., by bioturbation.  In this case, we should sum 
all of the signal, which we assume to have originated at the earliest time.

\end{enumerate}
In fact, as we now see, these scenarios lead to
similar predictions for $D$ and $\tsn$, as
they share key
aspects.  In both, the \fe60 was all produced
by the putative supernova at the explosion epoch.
Thus, the signal has decayed by the same factor
$e^{-\tsn/\tau}$, regardless of when the signal
arrived on earth and was deposited on the crust.
Thus, the signal in different crust layers should
not be given a different correction for decay.
Instead, the signals should all be added.
This is what we do for both scenarios, for the following
reasons. In
scenario 1 with contamination of the local interstellar medium, the
signal should have two components, an impulsive 
one received by the passage of the supernova blast through the solar 
system, and a continuous component derived from the solar system's
accretion of interstellar material enriched by its passage.  Both components
are signal, and may not be well-resolved,
depending on the relative strength, amplitude, and timescales
of each.  Thus, we sum the signals in all layers and take this to
be a rough estimate of the fraction of material deposited
on the earth. In
scenario 2 we posit that the signal is mixed across
layers.
Thus, even if we assume that the deposition
of material on the earth is only impulsive, the signal is smeared.  To
recover the original signal, we must again sum over the layers.

Thus, in both cases we take the signal to be
\beq
\label{eq:sum_signal}
N_{60}^{\rm tot} 
   = \sum_\ell N_{60}^{\oplus}(\Delta t_\ell) 
   \sim 5 \times 10^{6} \ {\rm atoms} \ {\rm cm}^{-2}
\eeq
Since only the \fe60 is taken as signal, we need to know one
of $D$ or $\tsn$ to get the other.  
An accurate estimate
of either is impossible with the available data, 
but different assumptions enable us to make non-trivial
limits and estimates, as seen in Fig.~1 and described in
the following paragraphs.

The data of Knie et al. in each time period are shown in Fig.~1 together
with their statistical error bars. As we have argued above,
the sum $N_i^{\rm tot}$ is the appropriate measure of the
total signal, and this is plotted as a vertical error bar for
\fe60 and \mn53 on the right-hand sides of the two panels.
The dashed lines also indicate the ranges favored for the
possible signals.

In terms of our fiducial values, \pref{eq:D_SN} now reads
\beq
D  =  29 \ e^{-\tsn/2 \tau_{60}} \ \psec  
\eeq
which we may use to derive constraints on the
supernova distance and epoch.
The maximum possible distance comes if the 
supernova happened ``yesterday'',
which is possible only in
the mixing scenario.  In this case, $\tsn/\tau_{60} \sim 0$,
and we have a maximum distance of 
\beq
D \la D_{\rm max} = 30 \ \psec 
\eeq
Interestingly, this distance happens to lie just within
the Ellis, Fields, \& Schramm \pcite{efs} estimate of the maximum distance 
at which a supernova might deposit its ejecta. 
Whilst the precision of the numerical agreement is
accidental, and subject to the numerous uncertainties
we have described, it is both amusing and intriguing that the
two are so close. The upper (lower) solid curve in the top panel
of Fig.~1 illustrates the signal expected from a
supernova exploding at a distance of 10~(30)~pc as a function
of the time at which it exploded. As described in \S2,
the astrophysical predictions have been corrected for the
reduced uptake via (\ref{eq:reduced}) with $f_{60} = 0.01$.
We note that this distance constraint comes about
by combining disparate information about supernova ejecta
and observed surface densities.  We find it both remarkable and
encouraging that these numbers combine to give a distance
limit that is not only of the right order of magnitude, 
but indeed jibes neatly with 
upper limit suggested by Ellis, Fields, \& Schramm \pcite{efs}.

A different consideration leads to another constraint
which limits the epoch of the
blast in both scenarios.  Since the putative supernova
apparently did not cause a 
catastrophic mass extinction, we require the
distance to be larger than the Ellis \& Schramm \pcite{es}
maximum killer radius: $D > 10 \ \psec$, which gives
\beq
\label{eq:nokill}
\tsn \le 4.6 \ \myr \ \bp
\eeq
as seen in Fig.~1.
Thus, even without identifying a crust layer with the epoch
of the \fe60 deposition, we can already limit
the distance to $10 \ \psec \la D \la 30 \ \psec$ and the epoch to 
$\tsn \la 5 \ \myr \ \bp$.  
The time constraints are not surprising, given that 
the very existence of the \fe60 essentially demands that
one place the putative supernova event within
the past few \fe60 lifetimes.

The result \pref{eq:nokill} 
raises an issue of self-consistency, since
$\tsn$ is so small as to be inconsistent with the
age of the lowest (i.e., oldest) crust layer,
$5.9-13.4 \myr \ \bp$.  This poses a problem in scenario 1,
which predicts that no signal should appear before the
explosion.  One can resolve this issue in two ways.
On the one hand, we can weaken the $\tsn$ limits
by taking a more conservative $2-\sigma$ lower limit
on the summed signal.
In this case, the age constraint rises to 5.9 Myr $\bp$,
just at the limit of the lowest layer's age.  On the other hand,
we note that only two events were detected in the lowest layer,
with a possible instrumental background
of one event, though Knie et al.\ argue that both events might be real
and thus have not made any subtraction.
If one regards the events in the oldest layers as background,
one should remove this layer's contribution to the
sum \pref{eq:sum_signal}.  In this case, the
time constraint to now gives a consistent
limit of $\tsn \le 5.4$ Gyr $\bp$.

We have so far used only the \fe60 data, assuming that
the observed \mn53 is background, but this assumption is
subject to challenge.  Whilst the detected flux is roughly consistent
with the expected background, our estimate is sufficiently
uncertain that it is worth examining the alternative.
Indeed, self-consistency demands that we estimate
the expected \mn53 signal, which
may be obtained directly from \mn53/\fe60 ratio.
The vertical error bar and dashed lines in the lower panel
of Fig.~1 represent the sum of the \mn53 data. The
curves in this panel represent the astrophysical
predictions for supernovae at distances of 10 or 30 pc as
before, assuming $f_{53} = 0.05$ and \mn53/\fe60 = 20.
As noted above, this ratio is unfortunately uncertain,
but we expect $\mn53/\fe60 \la 20$.
If the true value lies at the high end of this range, 
i.e., if $m_{\rm SN} \simeq 20 \msol$, then the observed \mn53 might
also be signal, as seen in the lower panel
of Fig.~1. In this case, the \mn53 signal
can no longer be used to estimate the reduced Mn uptake,
so we lose our ability to estimate $f_{53}$ and $f_{60}$.
On the other hand, if we interpret both radioisotopes as signal, we can 
derive the
supernova epoch.  Taking the net \mn53 surface density as pure signal
and using \pref{eq:t_SN}, we have
\beq
\label{eq:two_signals}
\tsn = 4.3 \myr \ \bp
\eeq
as also seen in Fig.~1, and
in good agreement with the range estimated above.
In terms of the (now unknown) reduced uptake for 
\mn53, this would imply a supernova distance of 
$D = 48 f_{53}^{1/2} \psec$, or $D \la 50 \psec$ for any
$f_{53} < 1$. 

It is already a new statement about supernova nucleosynthesis
if the \fe60 in the ferromanganese crust data indeed has a supernova
origin.
This isotope has long been predicted to come from supernovae.
In recent calculations, Woosley \& Weaver \pcite{ww} note that \fe60 is 
made in presupernova via $s$-process He burning,
and explosively at the base of the O and Si burning shells.
Observationally, there is meteoric evidence 
that live \fe60 was
present in the protosolar nebula, 
perhaps due to a supernova explosion soon before the formation of the 
solar system (Shukolyukov \& Lugmair \cite{sl92,sl93}).
Furthermore, \fe60 has received particular attention because its
decay through ${}^{60}$Co to ${}^{60}$Ni
is accompanied by the emission of 1.17 and 1.33 MeV $\gamma$ rays,
making it a target for search by $\gamma$-ray telescopes.
Timmes et al.\ \pcite{timmes} noted that the expected
$\gamma$-ray signal should
spatially trace that of ${}^{26}$Al.
They calculated the flux levels, and found them to be
just below the sensitivity of 
the Compton Gamma-Ray Observatory.  
However, \fe60 should be visible by
the upcoming INTEGRAL $\gamma$-ray satellite.  Until
seen in $\gamma$ rays, the data discussed here
could be the strongest available indication of
a supernova origin for \fe60.

The observation of additional
radioisotopes (Ellis, Fields, \& Schramm \cite{efs})
would not only help constrain the distance, 
but would also allow one
to use the various radioisotope measurements as
a telescope which provides information about the
supernova nucleosynthesis processes.
Therefore, we urge more searches, both of 
ferromanganese crusts like the one
reported (for repeatability and confirmation that the
effect is global), as well as other materials and
particularly other radioisotopes which would help confirm
the extraterrestrial origin and add to the constraints
on the timing, distance, and mass of the putative explosion.
Candidate species are those which are expected to be
made copiously in supernovae and have lifetimes comparable
to the $\sim 10 \myr$ timescale considered.  
Promising candidates include \be10,
\i129, and \sm146.  
Note that \be10  has a half-life
$\t12 = 1.51 \myr$ which is equal to that of
\fe60, within errors.  Thus, the \be10/\fe60 ratio
due to a nearby supernova should remain constant
over time, providing a consistency check.
Also, \be10 will have contributions from enhanced 
cosmogenic production as well as any possible supernova origin.
The other isotopes of note,
\i129 ($\t12 = 15.7 \myr$) and \sm146 ($\t12 = 10.3 \myr$),
have longer lifetimes, and thus can probe earlier epochs.
This could again allow for a cross-check: 
if \i129 and \sm146 are enhanced along with \fe60,
they should drop off in material which dates prior to the
explosion event.

To get a feel for the likelihood of the putative nearby supernova, 
we estimate the expected rate for an explosion
occurring with a given distance.
Following Shklovsky \pcite{shkl},
the average rate $\lambda$ of supernovae within a distance $D$ 
is just the total Galactic 
rate ${\cal R}_{\rm SN}$, times the volume fraction:
$\lambda = (4 D^3/3 R^2 h) \, {\cal R}_{\rm SN}$, where
$R$ is the disk radius and $h$ the scale height.
Using $R = 20$ kpc, $h$ = 100 pc, and 
the possibly optimistic
estimate ${\cal R}_{\rm SN} = 3 \times 10^{-2} \yr^{-1}$, 
we get
$\lambda \sim 1 \gyr^{-1} \, (D/10 \psec)^{3}$.
Thus an explosion at a distance of 30 pc 
should have a mean recurrence time of 100 Myr.
This is at least an order of magnitude larger
than the timescale suggested by the \fe60 data,
which suggests that either the supernova was
unusually recent, or that the rate has been grossly underestimated.
In this connection, we note that a recent and nearby supernova
remnant has been identified by Aschenbach \pcite{asch}. 
At an estimated distance
of about 200 pc, which is too distant to have any
effect of the type we discuss, but its age of about 700 yr 
(Iyudin et al.\ \cite{ti44}) does suggest that 
the rate of nearby supernovae could be higher than the above estimate.

Moreover, we note that the expected supernova
rate is enhanced
during the passage of the earth through spiral arms
(Shapley \cite{shap}; Hoyle \& Lyttleton \cite{hl}; 
Clark, McCrea, \& Stephenson \cite{cms}). This occurs every
$10^8$~yr or so, and it would be interesting to seek evidence
for any possible correlation with past extinctions. It seems that we
are now approaching or just entering the Orion arm,
so that an elevated supernova rate is possible.

\section{Impact on the Biosphere}

Potential implications of a nearby supernova explosion for
earth's biosphere have been considered by a number of authors
(Ruderman \cite{rude}; Ellis \& Schramm \cite{es};
Ellis, Fields, \& Schramm \cite{efs}), and
recent work has suggested that the most important effects might be induced
by cosmic rays. In particular, their possible role in destroying the
earth's ozone layer and opening the biosphere up to irradiation by
solar ultraviolet radiation has been emphasized (Ellis \& Schramm
\cite{es};
Ellis, Fields, \& Schramm \cite{efs}). The energetic
radiation from
supernovae contains two components: a neutral one
due to $\gamma$ rays, which has been estimated
to have a fluence
\beq
\phi_{\gamma} \sim 6.6 \times 10^5 \left ({ 10 \over D } \right )^2 {\rm
erg \, cm}^{-2}
\label{eq:neutralcr}
\eeq
for about a year, where here and subsequently $D$ is understood
to
be in units of pc, and a charged component whose flux has been estimated
to have a fluence
\beq
\phi_c \sim 7.4 \times 10^6 \left ({10 \over D} \right ) {\rm erg \ cm}^{-2}
\label{eq:chargedcr}
\eeq
for about $3 \, D^2$~yr, to be compared with the ambient flux of
$9 \times 10^4 \, {\rm erg \, cm}^{-2} {\rm yr}^{-1}$. 
We see that the ambient
flux would be doubled if $D \sim 30$~pc, with considerable
uncertainties, and could be considerably greater if $D \sim 20$~pc,
which cannot be excluded. These enhanced fluxes are
not thought likely to be directly dangerous to life.

However, it has been argued  (Ruderman \cite{rude}; Ellis \& Schramm \cite{es};
Ellis, Fields, \& Schramm \cite{efs}) that this ionizing
radiation should
produce NO in the stratosphere, making a contribution
\beq
y_{cr} \sim 88 \left ( {10 \over D } \right)^2
\label{crNO}
\eeq
to the NO abundance in parts per $10^9$. This is in turn estimated
to deplete the ozone abundance by a factor
\beq
F_O = { \sqrt{16 + 9 X^2} - 3 X \over 2 }
\label{lessO3}
\eeq
where $X = (3 + y_{cr})/3$ is the factor of enhancement in the abundance
of NO. In the case of a supernova at $30$~pc, we would estimate
$F_O \sim 0.33$. The
factor by which the penetrating flux of solar
ultraviolet radiation at the earth's surface is increased by
this ozone depletion is approximately $f^{F_O - 1}$, where $f$ is
the fraction of the solar ultraviolet flux that normally reaches the
surface. In the case of radiation with a wavelength of $2500$ \AA, which
is effective for killing {\it Escherichia Coli} bacteria and
producing erythema (sunburn), $f \sim 10^{-40}$ normally. We 
estimate that supernova at $30$~pc might increase this by some $27$
orders of magnitude, for a period measured in thousands of years. Clearly
these estimates are very unreliable, but they
serve as a warning that the effects of such a supernova may not
be negligible.

All the biosphere is dependent on photosynthesizing organisms at
the bottom of both the terrestrial and marine food chains.
Terrestrial photosynthesis is most effective for red light with a
wavelength $\sim 570$~nm, and we do not know of any detailed studies
how it might be impacted by enhanced solar ultraviolet radiation.
On the other hand, carotenoids in phytoplankton shift their most
sensitive wavelength towards the blue. This might provide a mechanism
for an amplification of the possible effect on marine ecology
relative to terrestrial ecology. The effect
of enhanced solar ultraviolet radiation
on marine photosynthesis by phytoplankton has in fact been studied in
connection with the ozone hole in the Antarctic, and a decline in
the rate of photosynthesis by phytoplankton exposed in plastic
bags has been demonstrated (Smith et al.\ \cite{s92}), although this needs to
be understood in the context of other effects such as vertical
mixing and cloudiness (Neale, Davis, \& Cullen \cite{n98}). 

It is natural to ask at this point whether any significant extinction
events are known to have occurred within the past $10$~Myr or so during
which the apparent excess of \fe60 may have been deposited. Indeed,
there is evidence for a couple of minor extinctions: one during the
middle Miocene, about $13$~Myr ago, and one of lesser significance
during the Pliocene, about $3$~Myr ago (Sepkoski \cite{s86}). 
Impacts on marine animal
families near the bottom of the food chain
have been noted, including zooplankton such as tropical
foraminifers (which
eat phytoplankton), bivalves, gastropods and echinoids (whose diets
include plankton and debris). This is exactly the pattern that
might be expected from a major insult to marine photosynthesis. 
It is interesting to note that the stability of phytoplankton
community structure over 200~kyr has been demonstrated using
deep-ocean sediments (Schubert et al.\ \cite{s98}), and it would
be valuable to extend such studies to longer periods. 
It would be fascinating to devise
a single experiment that could correlate directly possible isotope and 
phytoplankton signatures of a supernova event~\footnote{We
observe in passing that a weak correlation has been observed between
magnetic field reversals and mass extinctions (Raup \cite{raup}). 
We note that
an enhanced cosmic-ray flux is one consequence of such a reversal.}.

Many other possible causes of such an insult
should be considered, including volcanism and meteor impact(s),
and we would like to mention another possibility that could also be
linked to a nearby supernova explosion. A strong correlation has 
been observed 
(Friis-Christensen \& Lassen \cite{fcl91}; 
Svensmark \& Friis-Christensen \cite{sfc97}; 
Svensmark \cite{s97})
between solar activity (particularly the solar sunspot cycle) and
the earth's cloud cover. It is thought that cosmic rays may help seed
cloud formation, and it has been suggested that
the correlation with the sunspot cycle might 
be due to the known modulation of the cosmic-ray flux during the
solar cycle (Ney \cite{n59}), 
which is due to variations in the solar wind.
Increased cloud cover is expected to reduce the earth's
surface temperature (Hartmann \cite{h93}), 
and it has been conjectured that the
lower global temperatures three centuries
ago might be related to the different level of sunspot activity
at that time~\footnote{Interestingly, William Herschel noted two centuries ago that
wheat prices were anticorrelated with sunspot numbers.}. We remark
that the large increase in the cosmic
ray flux that we estimate from a nearby supernova explosion
might seed a large increase in the cloud cover, possibly
triggering a ``cosmic-ray winter'' lasting for thousands of
years~\footnote{Increasing the cloud cover might also provide a mechanism
for reversals of the terrestrial magnetic field to trigger
global cooling, since field reduction during a reversal
could enable a higher cosmic-ray flux to reach the earth's
upper atmosphere.}. Indications from recent solar cycles are that
variations in the cosmic-ray flux by about 20\% might be correlated
with fractional changes of the cloud cover by about 3\%, corresponding in
turn to variations in the mean earth temperature by about $0.4$~K 
(Kirkby \cite{k98}). 
This is a very speculative possibility, since there are
considerable variations in the flux of cosmic rays at different
energies and latitudes, and their 
efficiency for seeding clouds is only guessed from a statistical
analysis. Also, the ensuing impact on the
environment would be very complex, though no obvious
mechanism that would enhance the effect on marine life
comes immediately to mind. However, we do at least note that
accelerator experiments to probe the possible seeding of clouds by
cosmic rays are now being considered (Kirkby \cite{k98}).

\section{Conclusions} 

We have discussed in this paper the implications of the
possible anomalous \fe60 signature of a nearby supernova explosion
reported recently by Knie et al.\ \pcite{fe60}. We re-emphasize that the
interpretation of this effect requires confirmation. This could be
addressed by searching for anomalies in other radioisotopes as
suggested here and in Ellis, Fields \& Schramm \pcite{efs}, by
checking that the \fe60 background is low as argued here and in
Knie et al. \pcite{fe60}, by
verifying that the \fe60 enhancement is global, and by checking
that the \fe60 signal is absent in earlier ferromanganese layers.

Nevertheless, if the signal is real, it is the first direct
evidence for the supernova production of \fe60, and may be used to 
constrain the possible distance and epoch of the putative supernova.
We find that a distance of about 30 pc is consistent with the
magnitude of the \fe60 signal, and that it should have occurred
about 4 Myr ago. If the supernova origin of the observed \fe60
is confirmed, this opens up a whole new era of supernova
studies using deep-ocean sediments as telescopes. We
draw particular attention to the interest of searching for
\be10, \i129, and \sm146 as well as \mn53 and \fe60.

Finally, we have been encouraged by the report of
Knie et al. to review the possible impact of a nearby
supernova explosion on the biosphere. In this connection, we
recall that a couple of mini-extinctions have been reported
within the past 10 Myr or so, during the Middle Miocene and Pliocene. It 
would be interesting to investigate 
whether either of these may be correlated with a supernova event.
We have noted in passing that an enhancement of the cosmic-ray
flux, such as that accompanying a nearby supernova explosion,
might increase the global cloud cover. It remains to be
seen whether this might induce significant climate change
such as, in an extreme case, a ``cosmic-ray winter''. If the
\fe60 signal reported by Knie et al. is confirmed, such
speculation would become more compelling.

\newpage

\acknowledgments
We acknowledge and remember with gratitude our late friend
and collaborator David Schramm: his enthusiasm and
insight made work on this subject particularly enjoyable.\\

We thank warmly
G\"{u}nther Korshinek for sharing unpublished
results with us, and for many informative discussions.
One of us (J.E.) also thanks Jasper Kirkby for interesting
conversations.

\newpage

\centerline{\bf FIGURE CAPTION}

\begin{enumerate}

\item
\label{fig:N_vs_t}
The upper (lower) panel compares the data of Knie et al. \pcite{fe60}
with our predictions for the
surface density of \fe60 (\mn53). The data in each time period
are plotted with vertical statistical error bars, and their sums
are indicated by the error bars on the right-hand sides of the panels
and the horizontal dashed lines. The upper (lower) solid curves
are the astrophysical predictions assuming a supernova explosion
at a distance of 10~(30)~pc.
To compare with the data, these predicted
global surface densities have been
corrected downward by reduced uptake 
factors \pref{eq:reduced}, using the
Knie et al.\ values $f_{60} = 1/100$ and
$f_{53} = 1/20$.

\end{enumerate}

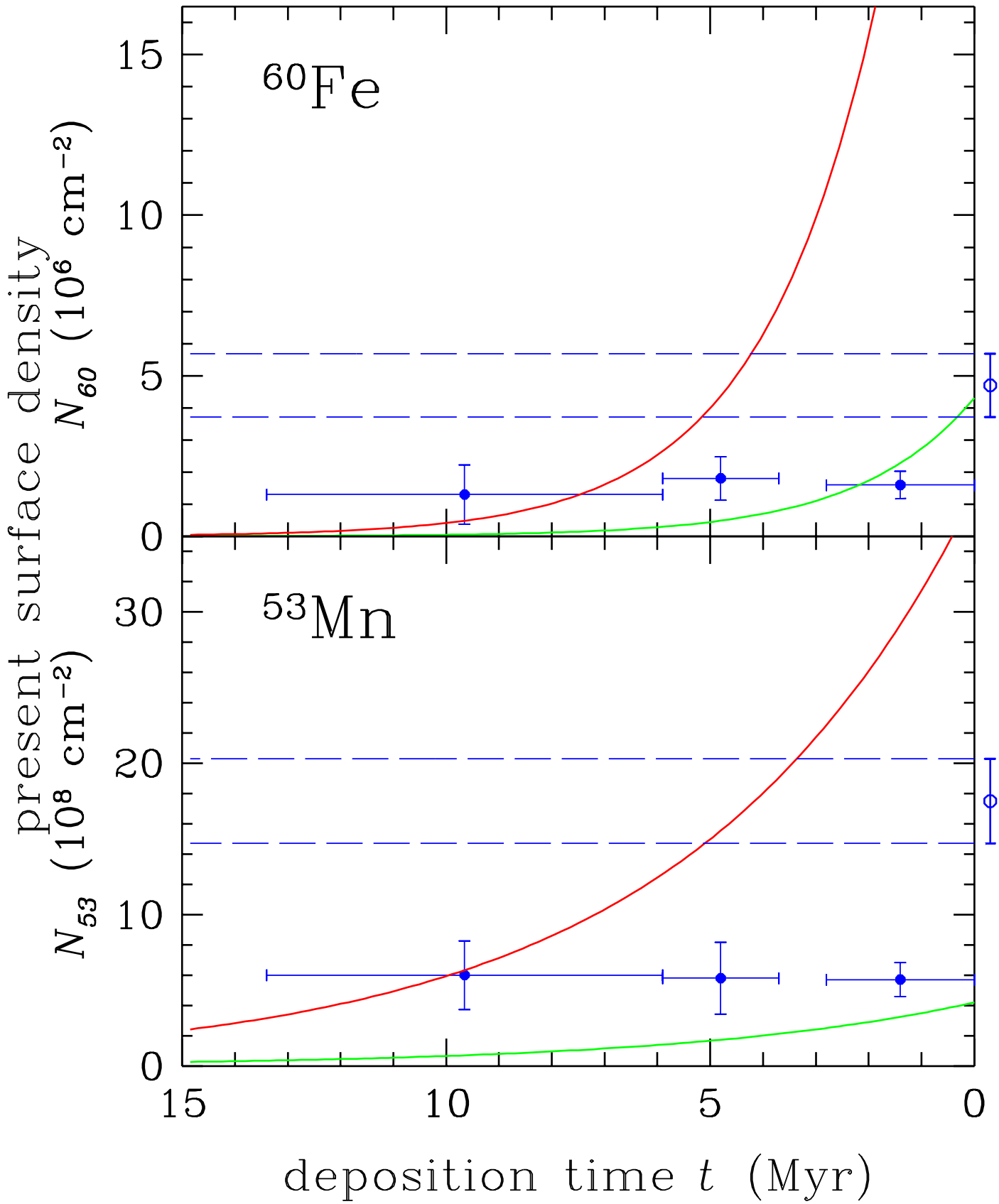
\end{document}